\documentclass[reprint,aps,prl,twocolumn,superscriptaddress]{revtex4-1}
\usepackage{graphicx}
\usepackage{dcolumn}
\usepackage{bm}
\usepackage{amsmath}
\usepackage[english]{babel}
\usepackage[utf8]{inputenc}
\begin{document}

\title{Ballistic supercurrent discretization and micrometer-long Josephson coupling in germanium}

\author{N. W. Hendrickx}
\email{n.w.hendrickx@tudelft.nl}
\author{M. L. V. Tagliaferri}
\author{M. Kouwenhoven}
\author{R. Li} 
\author{D. P. Franke}
\affiliation{QuTech and Kavli Institute of Nanoscience, Delft University of Technology, PO Box 5046, 2600 GA Delft, The Netherlands}
\author{A. Sammak}
\affiliation{QuTech and the Netherlands Organisation for Applied Scientific Research (TNO), Stieltjesweg 1, 2628 CK Delft, The Netherlands}
%\author{A. A. Golubov}
\author{A. Brinkman}
\affiliation{Faculty of Science and Technology and MESA+ Institute for Nanotechnology, University of Twente, 7500 AE Enschede, The Netherlands}
\author{G. Scappucci}
\author{M. Veldhorst}
\email{m.veldhorst@tudelft.nl}
\affiliation{QuTech and Kavli Institute of Nanoscience, Delft University of Technology, PO Box 5046, 2600 GA Delft, The Netherlands}

\date{\today}
\pacs{}

\begin{abstract}
We fabricate Josephson field-effect-transistors in germanium quantum wells contacted by superconducting aluminum and demonstrate supercurrents carried by holes that extend over junction lengths of several micrometers. In superconducting quantum point contacts we observe discretization of supercurrent, as well as Fabry-Perot resonances, demonstrating ballistic transport. The magnetic field dependence of the supercurrent follows a clear Fraunhofer-like pattern and Shapiro steps appear upon microwave irradiation. Multiple Andreev reflections give rise to conductance enhancement and evidence a transparent interface, confirmed by analyzing the excess current. These demonstrations of ballistic superconducting transport are promising for hybrid quantum technology in germanium.
\end{abstract}

\maketitle

Quantum information processing in the solid-state is being pursued using superconducting and semiconducting platforms \cite{Nakamura1999,Petta2005}. In both platforms, rudimentary quantum algorithms have already been demonstrated \cite{DiCarlo2009,Watson2018}. While decoherence is a central topic, advanced superconducting systems are now capable of entangling 10 qubits \cite{Song2017}. Spin qubits based on silicon (Si) and germanium (Ge), on the other hand, can be isotopically enriched to remove magnetic decoherence \cite{Itoh1993, Itoh2014}, resulting in extremely long coherence times \cite{Veldhorst2014, Sigillito2015}. Crucially, these qubits can be defined using conventional semiconductor technology. A hybrid approach may build upon the strengths of each platform motivating extensive research. Superconducting qubits with semiconductor elements have led to electric gate-tuneable superconducting qubits \cite{Larsen2015, Lange2015}, or gatemons, while spin qubits interfaced with superconducting resonators have reached the regime of strong spin-photon coupling \cite{Mi2018, Samkharadze2018, Landig2018}, an important step toward long-range entanglement.

Hybrid technology in condensed matter physics has even more surprises and can host exotic excitations. In particular, a topological phase transition may occur in superconductor-semiconductor systems in the presence of spin-orbit coupling and magnetism \cite{Sau2010, Alicea2010}. At the topological transition, excitations emerge that represent Majorana fermion states that can exhibit non-Abelian exchange statistics. Next to their fundamental interest, these states are argued to be excellent building blocks for quantum computation as they bear some topological protection against decoherence. Despite protection limited only to operations inside the Clifford group, coupling topological qubits to spin qubits may offer an effective pathway toward universal quantum computation \cite{Hoffman2016}. In addition, integrating topological systems to the spin qubit platform may enable the coupling of spatially separated spin qubits via topologically protected braiding operations \cite{Leijnse2011, Leijnse2012}.

Germanium has the potential to become an excellent material platform for the construction of these hybrid systems. It can be isotopically purified, thereby removing decoherence by nuclear spins \cite{Itoh1993}, and can host strong-spin orbit coupling \cite{Morrison2016}, in particular when the charge carriers are holes. In addition, mobilities reaching 1,500,000 cm$^2$/Vs have been reported \cite{Failla2016} and high-quality gate-defined quantum dots have been realized \cite{Hendrickx2018} in strained SiGe/Ge/SiGe heterostructures. Furthermore, electrically driven spin qubits have been constructed \cite{Watzinger2018}, single spins can be readout in single-shot mode \cite{Vukusic2018}, and germanium has several favorable properties for spin qubit operation \cite{Terrazos2018}, including a small effective mass and large energy splitting to excited states. Gate-tunable superconductivity has been reported in Ge/Si nanowires \cite{Xiang2006}, self-assembled Ge quantum dots \cite{Katsaros2010}, and more recently also in planar Ge structures \cite{Hendrickx2018}. A crucial question now is whether ballistic phase-coherent superconductivity can be induced in the germanium platform over the length scales required to harness the full capabilities of hybrid quantum technologies.

Here, we demonstrate gate-tunable Josephson supercurrents in planar germanium quantum wells with junction lengths $L$ up to 6 $\mu$m and find a characteristic decay length $\xi^*$ = 1.0 $\mu$m. In quantum point contacts we observe discretization of the supercurrent and conductance, demonstrating ballistic transport. From the excess current and multiple Andreev reflections we deduce an interface transparency $T$ between the leads and germanium of 0.6 and 0.7. Furthermore, we demonstrate the DC and AC Josephson effect in planar germanium via Fraunhofer-like patterns that arise in magnetic fields and Shapiro steps resulting from microwave irradiation.

\begin{figure}%
	\includegraphics[width=\linewidth]{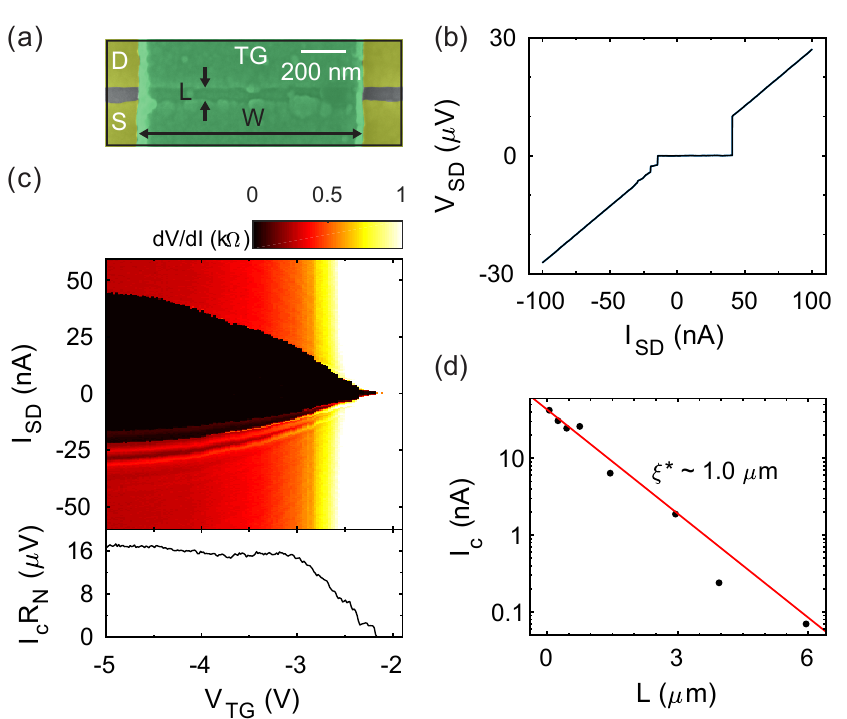}%
	\caption{(a) False-coloured SEM image of a planar Josephson junction device with width $W$ = 1 $\mu$m and length $L$ = 50 nm. The top gate $TG$ is used to induce a 2DHG in the strained germanium heterostructure contacted to a superconducting Al source $S$ and drain $D$. (b) IV-curve of the $L$ = 50 nm junction, showing a clear supercurrent with $I_C$ = 43 nA. (c, top panel) Colour plot of the differential resistance of the junction $dV/dI$ as a function of the bias current $I_{SD}$ and top-gate voltage $V_{TG}$. (c, bottom panel) $I_CR_N$ product as a function of $V_{TG}$. (d) Length dependence of the supercurrent. A purely exponential decay is observed over the entire range, with a decay length of 1.0 $\mu$m.}
\label{fig:JJ}
\end{figure}

Figure \ref{fig:JJ}(a) shows a scanning electron microscope (SEM) image of a germanium Josephson field effect transistor (JoFET). The strained SiGe/Ge/SiGe heterostructure is grown by reduced pressure chemical vapor deposition (RP-CVD)\cite{Hendrickx2018}. Superconducting leads are defined by thermal evaporation of Al after electron beam lithography and local etching of the Si capping layer. The top gate is fabricated by depositing a titanium (Ti) /  palladium (Pd) layer on top of an aluminum oxide (Al$_{2}$O$_{3}$) dielectric layer grown by atomic layer deposition (ALD) at $300 \ ^{\circ}$C. The JoFETs are fabricated with a junction length $L$ between 50 nm and 6 $\mu$m. The aluminum (Al) contacts are 1 $\mu$m wider than the width of the top gate $TG$, to ensure that the superconducting contact extends beyond the edges of the junction. 

A two-dimensional hole gas (2DHG) is formed by applying a negative gate voltage to the top gate and a clear supercurrent becomes apparent in the $IV$-curve. Figure \ref{fig:JJ}(b) shows a typical trace, where we find a critical current $I_C$ = 43 nA. The critical current can be tuned by changing the hole density using the top gate, as shown in the top panel of Fig. \ref{fig:JJ}(c). We find characteristic voltages $I_CR_N$ up to 17 $\mu$V, see bottom panel Fig. \ref{fig:JJ}(c), by taking the critical current and measuring the normal state resistance $R_N$ at high bias. Figure \ref{fig:JJ}(d) shows the length dependence of $I_C$, revealing supercurrents with a purely exponential decay length $\xi^*$ = 1.0 $\mu$m, extending over remarkably long length scales of several micrometers. For the junction with length $L$ = 6.0 $\mu$m, we measure $I_C$ = 70 pA. In comparison, supercurrents in semiconductors were reported with lengths up to 1.5 $\mu$m in graphene \cite{Calado2015}, 1.6 $\mu$m in GaAs \cite{Zhong2015}, 2 $\mu$m in InAs/GaSb \cite{Wenlong2014}, and 3.5 $\mu$m in Bi$_2$Te$_3$ \cite{Fanming2012}.  

\begin{figure}%
  \includegraphics[width=\columnwidth]{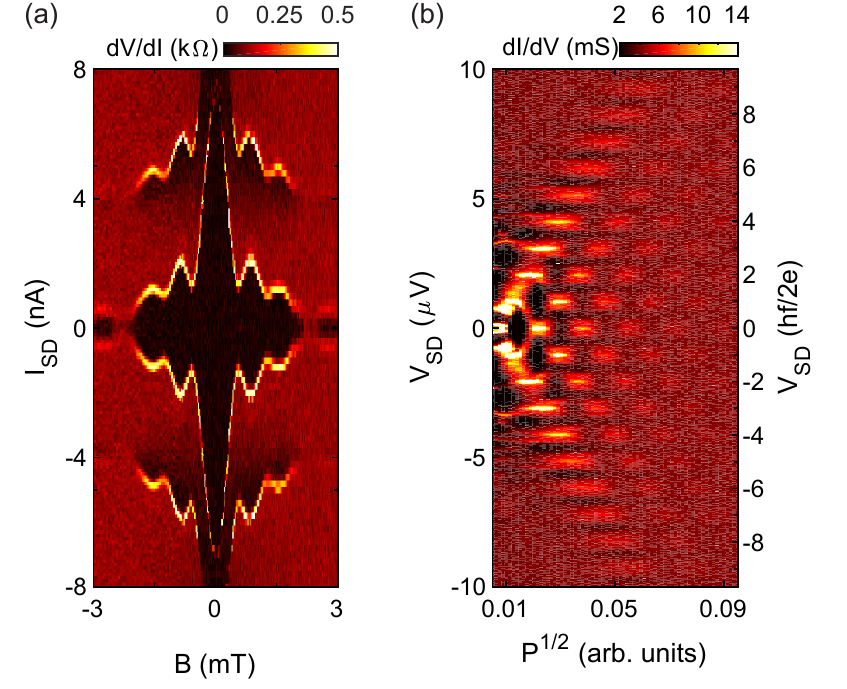}%
 \caption{(a)  Colour map of the differential resistance, $dV/dI$, of a junction with width $W=1.5\ \mu$m and length $L$ = 450 nm, as a function of magnetic field $B$ and bias voltage $V_{SD}$ at $V_{TG}$ = -5 V. A Fraunhofer-like modulation of the critical current as a function of $B$ can be observed, confirming the Josephson nature of the devices. (b) Differential conductance, $dI/dV$, of the junction with $L$ = 50 nm and $W$ = 1 $\mu$m as a function of bias current $I_{SD}$ and microwave excitation amplitude $P^{1/2}$, showing clear Shapiro steps at $V_{SD}$ = $nhf/2e$  $\approx n \times$ 1.03 $\mu$V, with excitation frequency $f$ = 500 MHz.}
  \label{fig:S&F}
\end{figure}

We test the Josephson nature of the supercurrent by performing phase sensitive experiments. Figure \ref{fig:S&F}(a) shows $I_C$ as a function of the external out-of-plane magnetic field $B$ for a junction with $L$ = 450 nm and $W$ = 1.5 $\mu$m. The modulation of the critical current follows a clear Fraunhofer-like pattern, demonstrating the DC Josephson effect. Based on the junction area one would expect a single flux quantum $\Phi_0$ through the junction at 3 mT, a magnetic field $\sim$5 times larger than we measured. The deviation suggests significant flux focusing. We also observe that the period decreases with increasing magnetic field and this could come from reduced flux focusing due to the Al layer leaving the Meissner state.

The AC Josephson effect gives rise to Shapiro steps in the presence of microwave irradiation. Clear plateaus are observed at $V_n$ = $nhf/2e$, see Fig. \ref{fig:JJ}(b), where we have applied an AC excitation with frequency $f$ = 500 MHz, such that $V_n$ = $n$ $\times$ 1.03 $\mu$V. We also observe steps positioned at $\delta$ = 0.22 $\mu$V on either side of a Shapiro step. In addition, in the absence of irradiation we observe multiple steps above $I_C$ that are linearly spaced in voltage, see Fig. \ref{fig:JJ}(a) for the first step. These steps are not yet understood, but may be of the same origin, are also independent of applied microwave frequency and observed in multiple junctions. We speculate these steps to originate from finite coupling of the junction to some cavity mode in or outside the device.

\begin{figure}%
  \includegraphics[width=\columnwidth]{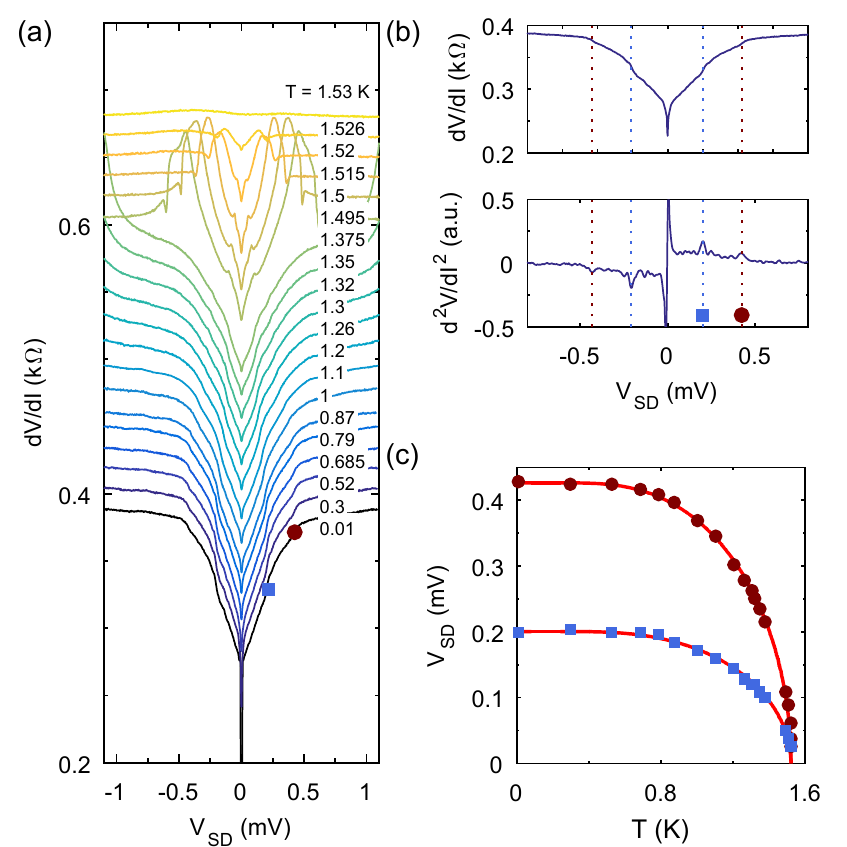}%
  \caption{a) Temperature dependence of the junction resistance at $V_{TG}$ = $-5.0$ V. The traces are successively offset for clarity. b) Resistance curve taken at 0.3 K (top panel) and its numerical derivative (bottom panel). The dotted lines show the position of the first and second MAR. Below 1.5 K their position is determined by the peaks in the numerical derivative, while above 1.5 K clear resistance peaks are visible. c) Temperature dependence of the first and second MAR features. Red lines are BCS fits scaled to match the MAR position at $T_{base} \approx$ 10 mK.}
  \label{fig:MAR}
\end{figure}

At higher bias voltage we observe changes in the conductance due to multiple Andreev reflection (MAR), see Fig. \ref{fig:MAR}. MAR appears at subgap voltages $V=2\Delta/me$, where $m$ is the number of Andreev reflections. To investigate the impact of MAR in more detail, we plot in Fig. \ref{fig:MAR}(b) the numerical derivative of the resistance and find that MAR causes clear peaks, measuring $\Delta$ = 0.2 mV. For tunnel contacts one would expect a resistance dip, while for transparent contacts Andreev reflection enhances the conductance and thus causes a resistance peak \cite{Averin1995}, as recently measured for an epitaxial aluminum/indium arsenide (Al/InAs) junction \cite{Kjaergaard2017}. By analyzing the excess current, see Fig. \ref{fig:JJ}(b) and details in the Supplementary Material, we find a junction transparency $T$ between 0.6 and 0.7, consistent with small MAR amplitude and resistance peaks. The MAR structure disappears when the temperature is above $T_C = 1.52$ K. We fit the data by scaling to the MAR positions at 10 mK \cite{Kjaergaard2017} and find good agreement with a pure BCS-like gap. The data suggests that $V_{m=2}/2-V_{m=1}\neq0$, which may be the result from the resistance not peaking at exactly $V=2\Delta/me$ \cite{Kjaergaard2017}. 

At even higher bias, we observe another resistance peak, see  Fig. \ref{fig:MAR}(a). The peak shifts to lower bias voltages with increasing temperature and disappears at $T_C$. At base temperature the peak is at $V_{SD}$ = 2 mV, which is at a bias voltage $\sim$10 times above the observed gap. A similar peak has been observed before \cite{Kjaergaard2017, Nguyen1992, Nguyen1994} and was attributed to non-equilibrium effects appearing in planar junctions where the high-mobility 2DHG extends underneath the superconducting contacts. Such an extended interface may increase the probability for Andreev reflection and could thereby be one of factors behind the observed transparent superconductor-semiconductor interfaces in planar structures.

\begin{figure}%
	\includegraphics[width=\linewidth]{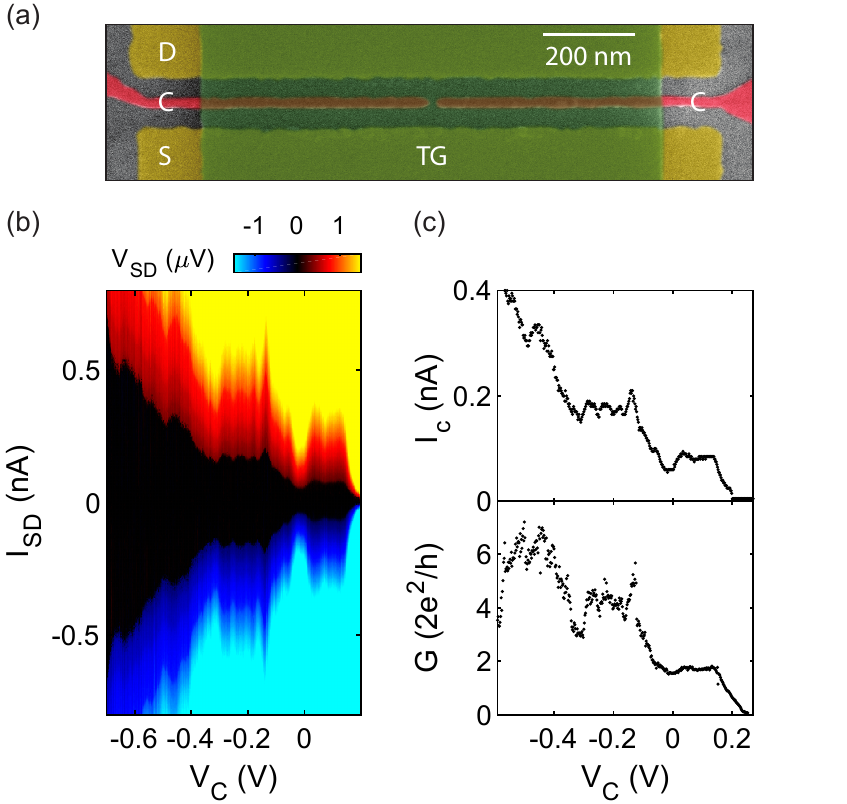}%
	\caption{(a) False-coloured SEM image of the superconducting QPC. A set of constriction gates $C$ are added to confine the number of transport channels through the junction. We set $V_{TG}$ = -2.8 V. (b) By tuning the constriction gate voltage $V_{C}$ we observe plateaus in the critical current due to a discrete number of modes in the QPC. We also observe oscillations in the conductance and current, which are most likely due to Fabry-Perot interference. (c) Discretization of critical current and conductance, demonstrating the ballistic nature of the superconducting device. Andreev reflection causes an enhanced conductance and steps exceeding $2e^2/h$.}
\label{fig:QPC}
\end{figure}

An important aspect for hybrid devices is whether transport can occur through individual channels. Quantum point contacts (QPCs) form an excellent playground to study the quantized nature of charge and were recently used to measure the strong $g$-factor anisotropy of heavy holes in strained SiGe/Ge/SiGe heterostructures \cite{Mizokuchi2018}. Here we focus on superconducting QPCs, predicted to give rise to supercurrent discretization \cite{Beenakker1991, Furusaki1991} with each mode in the QPC contributing $I_C = e \Delta/\hbar$ in the short and low temperature limit. Signatures of discrete supercurrents have been observed in InAs heterostructures \cite{Takayanagi1995, Bauch2005, Irie2014} and Si/Ge nanowires \cite{Xiang2006}. To study this in planar germanium, we have fabricated devices with a constriction close to the superconducting interfaces, see Fig. \ref{fig:QPC}a. Figure \ref{fig:QPC}b and \ref{fig:QPC}c show the transport characteristics. We tune the constriction gates $C$ to be more positive than the top gate $TG$. Upon increasing the voltage $V_C$ we see that first the transport underneath the constriction gates is turned off and for large enough positive voltages also the current through the constriction vanishes. In the intermediate regime, superconducting transport is carried by discrete modes, yielding clear discretization of the supercurrent and conductance. The conductance is measured at subgap voltage at a bias current of 1 nA. Owing to the conductance doubling of Andreev reflection, the conductance can raise in steps with amplitude larger than 2$e^2/h$ and we measure step heights $G_0$ = 3.4 $e^2/h$. The supercurrent increases in steps $I_{C0}$ = 85 pA. On top of the supercurrent discretization, we also observe regular conductance oscillations. We ascribe these to Fabry-Perot interference. Finite scattering at the constriction interfaces most likely cause the Fabry-Perot oscillations and may also cause the reduced QPC $I_CR_N$ = 1.1 $\mu V$ as compared to the JoFET $I_CR_N$ = 17 $\mu V$, Fig. \ref{fig:JJ}(c). 

The observation of a relatively low $I_CR_N$ product provides room for further investigation and possible optimization. Research could focus on epitaxial interfaces, although we already find rather high transparency. It may also be that transport is in the long-junction limit, even for the shortest junctions, and in this regime $I_CR_N$ is reduced. This would also explain the pure exponential length dependence of the critical current. A possible origin could be an extended interface, as speculated from the observed resistance peak at high bias, which would effectively increase the junction length. Alternatively, the coherence length could be unexpectedly short, for example due to transport carried by heavy holes with spin $J = 3/2$, strong spin-orbit coupling, and very anisotropic $g$-factors. While these speculations provide avenues for future research, the experimentally measured $I_CR_N$ = 17 $\mu$V already exceeds significantly the thermal energy at base temperature and clearly demonstrates proximity superconductivity in planar germanium. 

The gate-tunable Josephson supercurrent ranging over micrometer length-scales provides great opportunities for hybrid superconductor-semiconductor devices. Single particle transport as demonstrated in the superconducting quantum point contact provides further scope for experiments requiring individual modes. Planar gate-tunable superconducting qubits are within reach and could be coupled to nuclear spin-free spin qubits fabricated on the same platform. Topological qubits may require further development such as the observation of a hard gap, but could profit from the large $g$-factor of heavy holes and from the low disorder environment found in our systems. Germanium, the material that has led to the first transistor but was consequently surpassed by silicon in the information age, may thus strike back in the quantum information era.

\section*{Acknowledgements}
We thank William Lawrie for valuable discussions. MV acknowledges support through a FOM Projectruimte of the Foundation for Fundamental Research on Matter (FOM), associated with the Netherlands Organisation for Scientific Research (NWO).

\bibliography{references}

\begin{thebibliography}
\bibliographystyle{}
%% intro super-semi
\bibitem{Nakamura1999}{Y. Nakamura, Yu. A. Pashkin, and J. S. Tsai, Nature \textbf{398}, 786-788 (1999).}
\bibitem{Petta2005}{J. R. Petta, A. C. Johnson, J. M. Taylor, E. A. Laird, A. Yacoby, M. D. Lukin, C. M. Marcus, M. P. Hanson, and A. C. Gossard, Science \textbf{30}, 2180-2184 (2005).}
\bibitem{DiCarlo2009}{L. DiCarlo, J. M. Chow, J. M. Gambetta, Lev S. Bishop, B. R. Johnson, D. I. Schuster, J. Majer, A. Blais, L. Frunzio, S. M. Girvin, and R. J. Schoelkopf, Nature \textbf{460}, 240-244 (2009).}
\bibitem{Watson2018}{T. F. Watson, S. G. J. Philips, E. Kawakami, D. R. Ward, P. Scarlino, M. Veldhorst, D. E. Savage, M. G. Lagally, Mark Friesen, S. N. Coppersmith, M. A. Eriksson, and L. M. K. Vandersypen, Nature \textbf{555}, 633-637 (2018).} 
\bibitem{Song2017}{Chao Song, Kai Xu, Wuxin Liu, Chuiping Yang, Shi-Biao Zheng, Hui Deng, Qiwei Xie, Keqiang Huang, Qiujiang Guo, Libo Zhang, Pengfei Zhang, Da Xu, Dongning Zheng, Xiaobo Zhu, H. Wang, Y.-A. Chen, C.-Y. Lu, Siyuan Han, and J.-W. Pan, Phys. Rev. Lett. \textbf{119}, 180511 (2017).}
\bibitem{Itoh1993}{K. M. Itoh, W. L. Hansen, E. E. Haller, J. W. Farmer, V. I. Ozhogin, A. Rudnev, and A. Tikhomirov, J. Mater. Res. \textbf{8}, 1341 (1993).}
\bibitem{Itoh2014}{K.M. Itoh and H. Watanabe, MRS Commun. \textbf{4}, 143 (2014).}
\bibitem{Veldhorst2014}{M. Veldhorst, J. C. C. Hwang, C. H. Yang, A. W. Leenstra, B. de Ronde, J. P. Dehollain, J. T. Muhonen, F. E. Hudson, K. M. Itoh, A. Morello, and A. S. Dzurak, Nat. Nano. \textbf{9}, 981 (2014).}
\bibitem{Sigillito2015}{A. J. Sigillito, R. M. Jock, A. M. Tyryshkin, J. W. Beeman, E. E. Haller, K. M. Itoh, and S. A. Lyon, Phys. Rev. Lett. \textbf{115}, 247601 (2015).}
\bibitem{Larsen2015}{T. W. Larsen, K. D. Petersson, F. Kuemmeth, T. S. Jespersen, P. Krogstrup, J. Nygård, and C. M. Marcus, Phys. Rev. Lett. \textbf{115}, 127001 (2015).}
\bibitem{Lange2015}{G. de Lange, B. van Heck, A. Bruno, D. J. van Woerkom, A. Geresdi, S. R. Plissard, E. P. A. M. Bakkers, A. R. Akhmerov, and L. DiCarlo, Phys. Rev. Lett. \textbf{115}, 127002 (2015).}
\bibitem{Mi2018}{X. Mi, M. Benito, S. Putz, D. M. Zajac, J. M. Taylor, Guido Burkard, and J. R. Petta, Nature \textbf{555}, 599-603 (2018).}
\bibitem{Samkharadze2018}{N. Samkharadze, G. Zheng, N. Kalhor, D. Brousse, A. Sammak, U. C. Mendes, A. Blais, G. Scappucci, and L. M. K. Vandersypen, Science, dio:10.1126/science.aar4054 (2018).}
\bibitem{Landig2018}{A.J. Landig, J.V. Koski, P. Scarlino, U.C. Mendes, A. Blais, C. Reichl, W. Wegscheider, A. Wallraff, K. Ensslin, and T. Ihn, Nature (2018).}

%% topological qubits
\bibitem{Sau2010}{J. D. Sau, R. M. Lutchyn, S. Tewari, and S. Das Sarma, Phys. Rev. Lett. \textbf{104}, 040502 (2010).}
\bibitem{Alicea2010}{J. Alicea, Phys. Rev. B \textbf{81}, 125318 (2010).}
\bibitem{Hoffman2016}{S. Hoffman, C. Schrade, J. Klinovaja, and D. Loss, Phys. Rev. B \textbf{94}, 045316 (2016).}
\bibitem{Leijnse2011}{M. Leijnse, and K. Flensberg, Phys. Rev. Lett. \textbf{107}, 210502 (2011).}
\bibitem{Leijnse2012}{M. Leijnse, and K. Flensberg, Phys. Rev. B \textbf{86}, 104511 (2012).}

%% germanium paragraph
\bibitem{Morrison2016}{C. Morrison, J. Foronda, P. Wiśniewski, S. D. Rhead, D. R. Leadley, and M. Myronov, Thin Solid Films \textbf{602}, 84-89 (2016).}
\bibitem{Failla2016}{M. Failla, J. Keller, G. Scalari, C. Maissen, J. Faist, C. Reichl, W. Wegscheider, O. J. Newell, D. R. Leadley, M. Myronov, and J. Lloyd-Hughes, New. J. Phys. \textbf{18}, 113036 (2016).}
\bibitem{Hendrickx2018}{N. W. Hendrickx, D. P. Franke, A. Sammak, M. Kouwenhoven, D. Sabbagh, L. Yeoh, R. Li, M. L. V. Tagliaferri, M. Virgilio, G. Capellini, G. Scappucci, and M. Veldhorst, Nat. Comm. \textbf{9}, 2835 (2018).}
\bibitem{Watzinger2018}{H. Watzinger, J. Kukučka, L. Vukušić, F. Gao, T. Wang, F. Schäffler, J. Zhang, and G. Katsaros, arXiv:1802.00395.}
\bibitem{Vukusic2018}{L. Vukušić, J. Kukučka, H. Watzinger, J. M. Milem, F. Schäffler, and G. Katsaros, arXiv:1803.01775}
\bibitem{Terrazos2018}{L. A. Terrazos, E. Marcellina, S. N. Coppersmith, M. Friesen, A. R. Hamilton, X. Hu, B. Koiller, A. L. Saraiva, D. Culcer, and R. B. Capaz, arXiv:1803.10320.}
\bibitem{Xiang2006}{J. Xiang, A. Vidan, M. Tinkham, R.M. Westervelt, and C.M. Lieber, Nature Nano. \textbf{1}, 208-213 (2006).
\bibitem{Katsaros2010}{G. Katsaros, P. Spathis, M. Stoffel, F. Fournel, M. Mongillo, V. Bouchiat, F. Lefloch, A. Rastelli, O.G. Schmidt, and S. De Franceschi, Nat. Nano. \textbf{5}, 458-464 (2010).}


%% Supercurrent length
\bibitem{Calado2015}{V. E. Calado, S. Goswami, G. Nanda, M. Diez, A. R. Akhmerov, K. Watanabe, T. Taniguchi, T.M. Klapwijk, and L. M. K. Vandersypen, Nat. Nano. \textbf{10}, 761-764 (2015).}
\bibitem{Zhong2015}{Z. Wan, A. Kazakov, M. J. Manfra, L. N. Pfeiffer, K. W. West, and L. P. Rokhinson, Nat. Commun. \textbf{6}, 7426 (2015).}
\bibitem{Wenlong2014}{W. Yu, Y. Jiang, C. Huan, X. Chen, Z. Jiang, S. D. Hawkins, J. F. Klem, and W. Pan, Appl. Phys. Lett. \textbf{105}, 192107 (2014).}
\bibitem{Fanming2012}{F. Qu, F. Yang, J. Shen, Y. Ding, J. Chen, Z. Ji, G. Liu, J. Fan, X. Jing, C. Yang, and L. Lu, Sci. Rep. \textbf{2}, 339 (2012).}

%% MAR paragraph
\bibitem{Averin1995}{D. Averin and A. Bardas, Phys. Rev. Lett. \textbf{75}, 1831 (1995).}
\bibitem{Kjaergaard2017}{M. Kjaergaard, H. J. Suominen, M. P. Nowak, A. R. Akhmerov, J. Shabani, C. J. Palmstrøm, F. Nichele, and C. M. Marcus, Phys. Rev. Appl. \textbf{7}, 034029 (2017).}
%\bibitem{Golubov2004}{A. A. Golubov, M. Yu. Kupriyanov, and E. Il’ichev, Rev. Mod. Phys. \textbf{76}, 411 (2004).}

%% Additional peak
\bibitem{Nguyen1992}{C. Nguyen, H. Kroemer, and E. L. Hu, Phys. Rev. Lett. \textbf{69}, 2847 (1992).}
\bibitem{Nguyen1994}{C. Nguyen, H. Kroemer, E. L. Hu, Appl. Phys. Lett. \textbf{65}, 103 (1994).}

%% QPC
\bibitem{Mizokuchi2018}{R. Mizokuchi, R. Maurand, F. Vigneau, M. Myronov, and S. De Franceschi, Nano Lett., dio:10.1021/acs.nanolett.8b01457 (2018).}
\bibitem{Beenakker1991}{C. W. J. Beenakker and H. van Houten, Phys. Rev. Lett. \textbf{66}, 3056 (1991).}
\bibitem{Furusaki1991}{A. Furusaki, H. Takayanagi, M. Tsukada, Phys. Rev. Lett. \textbf{67}, 132 (1991).}
\bibitem{Takayanagi1995}{H. Takayanagi, T. Akazaki, J. Nitta, Phys. Rev. Lett. \textbf{75}, 3533 (1995).}
\bibitem{Bauch2005}{T. Bauch, E. Hurfeld, V.M. Krasnov, P. Delsing, H. Takayanagi, and T. Akazaki, Phys. Rev. B \textbf{71}, 174502 (2005).}
\bibitem{Irie2014}{H. Irie, Y. Harada, H. Sugiyama, T. Akazaki, Phys. Rev. B \textbf{89}, 165415 (2014).}

}
\end{thebibliography}

\end{document}